\documentstyle[preprint,aps]{revtex}
\begin{document}

\begin{titlepage}
\draft
\title{Dirac Neutrino Masses in NCG
     \footnote
{This work is supported by the Scientific Project of the Department of
Science and Technology in China, the National Scientific Foundation
in China and the Doctoral Program of the Foundation Institute of Higher
Education.
}}
\author{{\ Li Changhui, Ding Haogang, Dai Jian, Song Xingchang}}
\address{{\ Department of Physics, Peking University, Beijing 100871,China}}
\maketitle

\begin{abstract}
Several models in NCG with mild changes to the standard model(SM)
are introduced to discuss the neutrino mass problem. We use two constraints, 
Poincar$\acute{e}$ duality and gauge anomaly free, to discuss the possibility 
of containing right-handed neutrinos in them. Our work shows that no model in
this paper, with each generation containing a right-handed neutrino, can satisfy 
these two constraints in the same time. So, to consist with neutrino oscillation 
experiment results, maybe fundamental changes to the present version of NCG are 
usually needed to include Dirac massive neutrinos.
\end{abstract}

\end{titlepage}

%********************************************

\section{\bf INTRODUCTION}

Recent years, several experiments suggest that neutrino oscillations might
exist\cite{ref1}, so it is interesting to discuss the neutrino-mass problem
in NCG . Since majorana mass can't exist in the present NCG framework, we
assume that the possible neutrino mass is Dirac mass and the oscillation's
origin comes from the mixture in the lepton mass matrix.

Noncommutative geometry (NCG) gives us new insights into the SM. In the past
ten years, it has developed several versions\cite{ref2,ref3,ref4,ref5}. In
this paper, we only use the Connes-Lott's new scheme(a real spectral triplet
in it)\cite{ref6,ref7,ref8}, which has achieved many successes in SM\cite
{ref4,ref9}. Because three generations of right-handed neutrinos can not
exist in it\cite{ref4}, we must consider other models beyond SM, such as 
\cite{ref10,ref11,ref12,ref13}, if it is needed to discuss the neutrino mass
problem. To our opinion, any new models in NCG must obey some physical and
geometrical principles. In this paper we use such two constraints: {\em %
gauge anomaly cancellation }(physical) and {\em Poincar}$\acute{e}${\em \
duality}{\bf \ }(geometrical) in examining several models. Some of these
models appeared before, for example in\cite{ref11} the author considered the
right-handed neutrinos without changing the algebra structure of SM, and
concluded that if the number of the right-handed neutrinos and u-type quarks
are different , the model might contain massive neutrinos (we will get
different conclusions, seeing the following). Other models give changes to
the algebra structure of SM, but they have not considered right-handed
neutrinos.We will do some change in model-building concerning the neutrino
mass problem in NCG..

This paper is organized as follows: we first give a concise description of
Connes-Lott's model for SM, then we introduce the {\em Poincar}$\acute{e}$%
{\em \ duality}{\bf \ }and {\em gauge anomaly cancellation }in NCG. In
section {\rm III}, we use these two constraints to check some models to find
whether they can contain right-handed neutrinos, finally a conclusion
follows. %*****************************************************

\section{NCG in Standard Model}

We refer to\cite{ref4} for a clear and thorough review of Connes-Lott's
version of NCG. Here we emphasize on some basic aspects of NCG which are
needed for future discussions.

The fundamental element of the Connes-Lott's model(CL) is a real even
spectral triple $\left( {\cal A},{\cal H},{\cal D}\right) $ with a chirality
operator $\gamma $ and an antilinear isometry operator $J$ . In which ${\cal %
H}$ is a Hilbert space expanded by fermions and their antiparticles, ${\cal A%
}$ is an associative involution algebra representing on ${\cal H}$, ${\cal D}
$ is a self-adjoint Dirac operator, $\gamma $ and $J$ are also expressed as
operators on ${\cal H}$. In fact the triple in SM\ is a product of two
triples: one encodes spacetime, the other concerns internal space. In SM
(with one generation for example): 
\begin{eqnarray}
{\cal A} &=&C^\infty (M,C)\otimes {\cal A}_F  \nonumber \\
{\cal H} &=&L^2(M,S)\otimes {\cal H}_F  \nonumber \\
{\cal D} &=&({\partial \mkern-9mu/}\otimes 1)\oplus (1\otimes {\cal D}_F) 
\nonumber
\end{eqnarray}

In this paper, discussing the possibility of introducing right-hand neutrino
in NCG model, we only concern the internal space, which is a finite spectral
triple sometime called as the finite-part $K-cycle$. Their definitions in SM
are:

${\cal A}_F=C\oplus H\oplus M_3\left( C\right) $, ($H$ is the algebra of
quaternion), ${\cal H}_F={\cal H}_F^{+}\oplus {\cal H}_F^{-}$.

${\cal H}_F^+$, ${\cal D}_F$, $J$, and $\gamma $ are defined as:

${\cal H}_F^{+}=\left( 
\begin{array}{c}
e_R \\ 
\left( 
\begin{array}{c}
\upsilon _L \\ 
e_L
\end{array}
\right) \\ 
u_R \\ 
d_R \\ 
\left( 
\begin{array}{c}
u_L \\ 
d_L
\end{array}
\right)
\end{array}
\right) $ , ${\cal D}_F=\left( 
\begin{array}{cccc}
0 & M & 0 & 0 \\ 
M^{*} & 0 & 0 & 0 \\ 
0 & 0 & 0 & \overline{M} \\ 
0 & 0 & \overline{M^{*}} & 0
\end{array}
\right) ,$ $J$ =$\left( 
\begin{array}{cccc}
0 & 0 & 1 & 0 \\ 
0 & 0 & 0 & 1 \\ 
1 & 0 & 0 & 0 \\ 
0 & 1 & 0 & 0
\end{array}
\right) \circ C$, $\gamma =\left( 
\begin{array}{cc}
1 & 0 \\ 
0 & -1
\end{array}
\right) $

${\cal H}_F^{+}$ represents the particle sector (we omit the analog
representation in ${\cal H}_F^{-}$). $M$ is mass matrix of fermions (in
multi-generation model it also contains the mixing angles such as CKM
matrix). The operator of $C$ in the definition of $J$ is${\cal \ }$the
complex conjugation operator. Chirality $\gamma $ is also an operator which
gives $1$ to right-handed particles and $-1$ to left-handed particles. A
noncommutative geometry must fulfill some axioms such as Poincar$\acute{e}$
duality , which are described in \cite{ref3,ref4,ref7}.

\section{Poincar$\acute{e}$ duality{\bf \ and anomaly cancellation
constraints}}

%*****************************************************************

\subsection{ Poincar\'{e} Duality}

There is a well known property for a close Riemannian manifold: {\it %
Poincar\'{e} duality}, which means that there exists an isomorphism between
the de Rham groups $H^p\left( V\right) $ and $H^{n-p}\left( V\right) $ ($n$
is the dimension of the manifold). Connes has put forth this property into
any real spectral triples\cite{ref3,ref7}. In order to show how it works, we
demonstrate the {\it Poincar\'{e} duality } in SM as an introduction. The
representation of ${\cal A}_F$ on ${\cal H}_F$ is :

for lepton sector:\ $\pi _l^{+}\left( \lambda ,q\right) =\left( 
\begin{array}{cc}
\lambda & 0 \\ 
0 & q
\end{array}
\right) \otimes 1_N,\ \pi _l^{-}\left( \lambda ,q\right) =\left( 
\begin{array}{lll}
\overline{\lambda } & 0 & 0 \\ 
0 & \overline{\lambda } & 0 \\ 
0 & 0 & \overline{\lambda }
\end{array}
\right) \otimes 1_N$

for quark sector:\ $\pi _q^{+}\left( \lambda ,q\right) =\left( 
\begin{array}{lll}
\overline{\lambda } & 0 & 0 \\ 
0 & \lambda & 0 \\ 
0 & 0 & q
\end{array}
\right) \otimes 1_3\otimes 1_N,\ \pi _q^{-}\left( \lambda ,q,m\right)
=\left( 
\begin{array}{llll}
m & 0 & 0 & 0 \\ 
0 & m & 0 & 0 \\ 
0 & 0 & m & 0 \\ 
0 & 0 & 0 & m
\end{array}
\right) \otimes 1_N$

in which $\lambda ,q,m $ are elements of $C,H,M_3(C)$ respectively. $\left(
+,-\right) $ denotes particle and anti-particle sector respectively. $N$
denotes the number of generations of fermions.

Define :

\begin{equation}
Q_{ij}=(p_{i,}p_j)=Tr(\gamma p_iJp_jJ^{\dagger })
\end{equation}

where $p_i$ is the minimal-rank hermitian projection of an algebra such as: $%
1_C$ for $C$, $1_H(I_2)$ for $H$, and the diagonal matrix $e=(1,0,0)$ for $%
M_3\left( C\right) $. We choose the same bases as \cite{ref4} in which $%
p_1=\left( -1_C\right) \oplus e,$ $p_2=1_C\oplus 1_H,$ $p_3=1_C$. Poincar$%
\acute{e}$ duality in NCG requires this matrix has non-vanishing
determinant, that is:

\begin{equation}
DetQ\neq 0
\end{equation}
In the SM the chirality and projections take the form:

\begin{eqnarray}
\gamma &\mapsto &\left( 1,-1,-1\right) ^N\oplus \left( 1,1,-1,-1\right)
^{3N}\oplus \left( 1,-1,-1\right) ^N\oplus \left( 1,1,-1,-1\right) ^{3N} 
\nonumber \\
p_1 &\mapsto &\left( -1,0,0\right) ^N\oplus \left( -1,-1,0,0\right)
^{3N}\oplus \left( -1,-1,-1\right) ^N\oplus \left( e,e,e,e\right) ^N 
\nonumber \\
p_2 &\mapsto &\left( 1,1,1\right) ^N\oplus \left( 1,1,1,1\right) ^{3N}\oplus
\left( 1,1,1\right) ^N\oplus \left( 0,0,0,0\right) ^{3N}  \nonumber \\
p_3 &\mapsto &\left( 1,0,0\right) ^N\oplus \left( 1,1,0,0\right) ^{3N}\oplus
\left( 1,1,1\right) ^N\oplus \left( 0,0,0,0\right) ^{3N}  \nonumber
\end{eqnarray}

From definition (1):

\[
Q=-2\left( 
\begin{array}{ccc}
N & 0 & 0 \\ 
0 & N & 0 \\ 
0 & 0 & -N
\end{array}
\right) 
\]

Its determinant is obviously nonzero so that the SM in NCG satisfies Poincar$%
\acute{e}$ duality requirement. But if we put $N$ generations right-handed
neutrinos into SM without any other changes, we will get a vanishing matrix $%
Q$ \cite{ref4}(that's why we need other models beyond SM). In the following,
(2) is used as the Poincar\'{e} duality constraint in model building. 
%*******************************************************

\subsection{Gauge Anomaly Cancellation}

Gauge anomaly free is required by the renormalizability in QFT. In ordinary
quantum field theory the anomaly is proportional to $tr[\gamma^5\lambda _a\left
\{ \lambda_{b,}\lambda _c\right\}]$ \cite{ref14}, where $\lambda _i$ are the
generators of the gauge group. In CL model, Gauge group is obtained from the
unitary elements of the algebra ${\cal A}_F$. The anomaly free condition in
NCG generally is\cite{ref10}:

\begin{equation}
Tr_p\left[ \gamma \left( \pi \left( x\right) +J\pi \left( x\right)
J^{\dagger }\right) ^3\right] =0
\end{equation}
$Tr_p$ is the trace on ${\cal H}_F$ restricted to particle sector ${\cal H}%
_F^{+}$ and $x$ is a unitary element of ${\cal A}_F$. Here we still use
standard model as an example: we refer \cite{ref10} to find details.

From the representations of $\pi \left( \lambda ,q,m\right) $, we get $\pi
\left( a,b,c\right) +J\pi \left( a,b,c\right) J^{\dagger }$ in particle
sector:

\[
\pi \left( a,b,c\right) +J\pi \left( a,b,c\right) J^{\dagger }=diag\left( 
\begin{array}{c}
2b \\ 
a+b1_2 \\ 
c+\left( ub-b\right) 1_3 \\ 
c+\left( ub+b\right) 1_3 \\ 
a\otimes 1_3+1_2\otimes c+ub\otimes 1_6
\end{array}
\right) \otimes 1_N 
\]
in which $i\left( a,b,c\right) \in su\left( 2\right) \oplus R\oplus su\left(
3\right) $. We has used the decomposition $u\left( 3\right) =u\left(
1\right) \oplus su\left( 3\right) $, then we write the Lie-Algebra of
unitaries of $M_3(C)$ in terms of $c+ub$ (in SM, only one $U(1)$ gauge boson
exists, so here is only $b$, no $b^{\prime }$, in models where there are
more than one $U(1)$ gauge bosons, the $b^{\prime }$ is needed ), $u$ is an
arbitrary real number.

From equation (3) we can get three equations (we only concern $U(1)$ gauge) :

\begin{eqnarray}
N\left( 8-2\right) +3N[ \left( u-1\right) ^3+\left( u+1\right)
^3-2u^3] &=&0 \\
N+3Nu &=&0 \\
N\left( u-1\right) +N\left( u+1\right) -2Nu &=&0
\end{eqnarray}

The result is: $u=-1/3$, which is the same result as it from the so-called
unimodularity condition\cite{ref4}:

\[
Tr_p\left[ \gamma \left( \pi \left( x\right) +J\pi \left( x\right)
J^{\dagger }\right) \right] =0 
\]

In the following discussions, we use (3) as the gauge anomaly cancellation
condition. %***********************************************************

\section{Right-handed Neutrino in Several Models}

We will begin to discuss several models beyond SM: Model 1 considers the
possibility of introducing right-hand neutrino in NCG in terms of changing
the fermion representation ${\cal {H}}_F$; model 2 is originally studied by
others\cite{ref10} to find whether there is another $U(1)$ gauge boson, in
which another $C$ algebra was put in; model 3 comes from the mathematics
consideration on quantum group, where quaternion algebra was changed to $%
M_2(C)$ ; model 4 is a combination of model 2 and 3. They all give mild
changes to the Connes-Lott's version of SM, and we will introduce
right-neutrinos in them. Our works show that the two constraints do not
permit all generations have right-handed neutrino in those models. We also
get a different conclusion in Model 1 from its previous conclusion in \cite
{ref11}. %**************************************************************

\subsection{Model {\rm 1}:}

In\cite{ref11}, Rich Schelp thought out a possible way to put right-handed
Dirac neutrinos into SM, which assumes there are $N_1$ massless generations
of right-handed neutrinos and $(N-N_1)$ massive ones. The representations
now are: (for those massless fermions, the representations on their
right-handed particle are all zero)

\[
\pi _{l1}^{+}(\lambda ,q)=\left( 
\begin{array}{ccc}
0 & 0 & 0 \\ 
0 & \lambda & 0 \\ 
0 & 0 & q
\end{array}
\right) \otimes 1_{N_1},\ \ \pi _{l2}^{+}(\lambda ,q)=\left( 
\begin{array}{lll}
\overline{\lambda } & 0 & 0 \\ 
0 & \lambda & 0 \\ 
0 & 0 & q
\end{array}
\right) \otimes 1_{N-N_1} 
\]

The same assumption is put for $u-type$ quarks except substituting $%
N_1\rightarrow N_2$(we omit the analog representation in antiparticle
sector). Then we can write down those $p_i.$ 
\begin{eqnarray}
\gamma  &\mapsto &\left( 1,1,-1,-1\right) ^N\oplus \left( 1,1,-1,-1\right)
^{3N}\oplus \left( 1,1,-1,-1\right) ^N\oplus \left( 1,1,-1,-1\right) ^{3N} 
\nonumber \\
p_1 &\mapsto &\left( 0,-1,0,0\right) ^{N_1}\oplus \left( -1,-1,0,0\right)
^{N-N_1}\oplus \left( 0,-1,0,0\right) ^{3N_2}\oplus \left( -1,-1,0,0\right)
^{3\left( N-N_2\right) }  \nonumber \\
&\oplus &\left( 0,-1,-1,-1\right) ^{N_1}\oplus \left( -1,-1,-1,-1\right)
^{N-N_1}\oplus \left( 0,e,e,e\right) ^{N_2}\oplus \left( e,e,e,e\right)
^{N-N_2}  \nonumber \\
p_2 &\mapsto &\left( 0,1,1,1\right) ^{N_1}\oplus \left( 1,1,1,1\right)
^{N-N_1}\oplus \left( 0,1,1,1\right) ^{3N_2}\oplus \left( 1,1,1,1\right)
^{3\left( N-N_2\right) }  \nonumber \\
&\oplus &\left( 0,1,1,1\right) ^{N_1}\oplus \left( 1,1,1,1\right)
^{N-N_1}\oplus \left( 0,0,0,0\right) ^{3N_2}\oplus \left( 0,0,0,0\right)
^{3\left( N-N_2\right) }  \nonumber \\
p_3 &\mapsto &\left( 0,1,0,0\right) ^{N_1}\oplus \left( 1,1,0,0\right)
^{N-N_1}\oplus \left( 0,1,0,0\right) ^{3N_2}\oplus \left( 1,1,0,0\right)
^{3\left( N-N_2\right) }  \nonumber \\
&\oplus &\left( 0,1,1,1\right) ^{N_1}\oplus \left( 1,1,1,1\right)
^{N-N_1}\oplus \left( 0,0,0,0\right) ^{3N_2}\oplus \left( 0,0,0,0\right)
^{3\left( N-N_2\right) }  \nonumber
\end{eqnarray}

From definition of (1) :

\[
Q=-2\left( 
\begin{array}{ccc}
N_1-N_2 & N-N_1+\frac 12N_2 & N-N_1+\frac 12N_2 \\ 
N-N_1+\frac 12N_2 & N_1 & N_1-N \\ 
N-N_1+\frac 12N_2 & N_1-N & N_1-2N
\end{array}
\right) 
\]

and it's determinant

\[
DetQ=8\left( N_1-N_2\right) N^2 
\]

Poincar$\acute{e}$ duality requires $N_1-N_2\neq 0.$ For the second
constraint, next to calculate:

\[
\pi \left( x\right) +J\pi \left( x\right) J^{\dagger }=diag\left( 
\begin{array}{c}
\left( 0\right) \otimes 1_{\left( N-N_1\right) } \\ 
2b\otimes 1_N \\ 
\left( a+b\right) \otimes 1_N \\ 
\left( -b+ub+c\right) \otimes 1_{\left( N-N_2\right) } \\ 
\left( b+ub+c\right) \otimes 1_N \\ 
\left( a+ub+c\right) \otimes 1_N
\end{array}
\right) 
\]

then from gauge anomaly cancellation (3), we get the following equations :

\begin{eqnarray}
N\left( 8-2\right) +3[ \left( N-N_2\right) \left( u-1\right) ^3+N\left(
u+1\right) ^3-2Nu^3] &=&0 \\
N+3Nu &=&0 \\
\left( N-N_2\right) \left( u-1\right) +N\left( u+1\right) -2Nu &=&0
\end{eqnarray}
the solution is $u=-1/3$, and $N_2=0$. $N_2$ is zero means that all the $%
u-type $ quarks have masses, which is different from the conclusion in \cite
{ref11}(in which only Poincar$\acute{e}$ duality is considered). On the
other hand, $N_1-N_2\neq 0$ , together with $N_2=0$, educes that if the
right-handed neutrino exists in this model, the number of generations of it
is less than $N$ ( in SM $N=3$ ). But it seems unnatural and the limited
experiment results up-to-now (if correct) do not support it (those
experiments need at least three massive neutrinos). 
%***********************************************************

\subsection{Model {\rm 2}: $U\left( 1\right) $ extension}

Another extension of standard model is discussed in paper \cite{ref10}, in
which the algebra ${\cal A}_F$ has two $C$ algebras $C$ and $C^{\prime }$.
Now we put $N$ generations right-handed neutrinos in this model, leaving the
weak and strong sectors unchanged as it has been done in \cite{ref10}. The
properties of NCG\cite{ref4}: $\left[ \pi (a),J\pi (b)J^{\dagger }\right]
=0\ ,\ \left[ \left[ D,\pi (a)\right] ,J\pi (b)J^{\dagger }\right] =0\ \
for\ a,b\in {\cal A}$, make the representation of algebra ${\cal A}_F$ on
antiparticle part of leptons is vectorial\cite{ref4} (all belong to algebra $%
C$ or $C^{\prime }$ ). Without losing generality we can assume they all
belong to the first $C$ algebra. Then the algebra and it's representation
are: (from now on, we use the Lie-Algebra directly instead of algebra ${\cal %
A}_F$ for convenience )

${\cal A}_F=C\oplus C^{\prime }\oplus H\oplus M_3(C);$

$\pi _R^{+}\left( b,b^{\prime }\right) =diag\left( \left( y_vb+y_v^{\prime
}b^{\prime }\right) I_N,\left( y_eb+y_e^{\prime }b^{\prime }\right)
I_N,\left( y_ub+y_u^{\prime }b^{\prime }\right) I_{3N},\left(
y_db+y_d^{\prime }b^{\prime }\right) I_{3N}\right) ;$

$\pi _L^{+}\left( a\right) =diag\left( aI_N,aI_{3N}\right) ;$

$\pi _R^{-}\left( b,b^{\prime },c\right) =diag\left( -bI_{2N},I_{2N}\otimes
\left( ubI_3+u^{\prime }b^{\prime }I_3+c\right) \right) ;$

$\pi _L^{-}\left( b,b^{\prime },c\right) =diag\left( -bI_{2N},I_{2N}\otimes
\left( ubI_3+u^{\prime }b^{\prime }I_3+c\right) \right) ;$

in which $y_i$ and $y_i^{\prime }\in \left\{ -1,0,1\right\} $, and for the
same $i,$ one and only one of $\left\{ y_i,y_i^{\prime }\right\} $ is zero 
\cite{ref10}( it is required by the representation of Algebras). So, $%
y_iy_i^\prime=0$. $u,u^{\prime }$ are two real numbers. Since there are two $%
C$ algebras now, we take the decomposition of $u(3)$ as $u\left( 3\right)
=u^{\prime }(1)\oplus u\left( 1\right) \oplus su\left( 3\right)$.

The minimal projection of algebra $C^{\prime }$ is $1_C^{\prime }$ , and we
define $p_4=1_C^{\prime }$. Other $p_i$s $(i=1,2,3)$ are the same as before,
but right-handed neutrinos change the chirality $\gamma $ to be the
following: 
\[
\gamma \mapsto \left( 1,1,-1,-1\right) ^N\oplus \left( 1,1,-1,-1\right)
^{3N}\oplus \left( 1,1,-1,-1\right) ^N\oplus \left( 1,1,-1,-1\right) ^{3N}
\]

In the same way we get the matrix: 
\[
Q=\left( 
\begin{array}{cccc}
2N\left( Y_e-Y_u\right) & -N\left( 2Y_e-Y_u\right) & -N\left( 2Y_e-Y_u\right)
& -N\left( Y_e^{\prime }-Y_u^{\prime }\right) \\ 
-N\left( 2Y_e-Y_u\right) & 2N\left( Y_e-2\right) & 2N\left( Y_e-1\right) & 
NY_e^{\prime } \\ 
-N\left( 2Y_e-Y_u\right) & 2N\left( Y_e-1\right) & 2NY_e & NY_e^{\prime } \\ 
-N\left( Y_e^{\prime }-Y_u^{\prime }\right) & NY_e^{\prime } & NY_e^{\prime }
& 0
\end{array}
\right) 
\]

and its determinant

\[
DetQ=4N^4\left( Y_e^{\prime }-Y_u^{\prime }\right) ^2 
\]

in which $Y_e=\left| y_v\right| +\left| y_e\right| $, $Y_u=\left| y_u\right|
+\left| y_d\right| $, $Y_e^{\prime }=\left| y_v^{\prime }\right| +\left|
y_e^{\prime }\right| $, $Y_u^{\prime }=\left| y_u^{\prime }\right| +\left|
y_d^{\prime }\right| $. {\em Poincar}$\acute{e}${\em \ duality} requires $%
Y_e^{\prime }\neq Y_u^{\prime }$. Again, we need to calculate: 
\[
\pi \left( x\right) +J\pi \left( x\right) J^{\dagger }=diag\left( 
\begin{array}{c}
\left( y_v+1\right) b+y_v^{\prime }b^{\prime } \\ 
\left( y_e+1\right) b+y_e^{\prime }b^{\prime } \\ 
a+b \\ 
\left( y_u+u\right) b+\left( y_u^{\prime }+u^{\prime }\right) b^{\prime }+c
\\ 
\left( y_d+u\right) b+\left( y_d^{\prime }+u^{\prime }\right) b^{\prime }+c
\\ 
a+ub+u^{\prime }b^{\prime }+c
\end{array}
\right) 
\]

From the gauge anomaly cancellation condition (3) we get: 
\begin{eqnarray}
\left( y_v+1\right) ^3+\left( y_e+1\right) ^3-2+3\left[ \left( y_u+u\right)
^3+\left( y_d+u\right) ^3-2u^3\right]  &=&0 \\
1+3u &=&0 \\
\left( y_u+u\right) +\left( y_d+u\right) -2u &=&0 \\
\left( y_v+1\right) ^2y_v^{\prime }+\left( y_e+1\right) ^2y_e^{\prime
}+3\left[ \left( y_u+u\right) ^2\left( y_u^{\prime }+u^{\prime }\right)
+\left( y_d+u\right) ^2\left( y_d^{\prime }+u^{\prime }\right)
-2u^2u^{\prime }\right]  &=&0 \\
{y_v^{\prime }}^3+{y_e^{\prime }}^3+3\left[ \left( y_u^{\prime }+u^{\prime
}\right) ^3+\left( y_d^{\prime }+u^{\prime }\right) ^3-2u{^{\prime }{}^3}%
\right]  &=&0 \\
u^{\prime } &=&0 \\
\left( y_u^{\prime }+u^{\prime }\right) +\left( y_d^{\prime }+u^{\prime
}\right) -2u^{\prime } &=&0 \\
\left( y_v+1\right) {y_v^{\prime }}^2+\left( y_e+1\right) {y_e^{\prime }}%
^2+3\left[ \left( y_u+u\right) \left( y_u^{\prime }+u^{\prime }\right)
^2+\left( y_d+u\right) \left( y_d^{\prime }+u^{\prime }\right) ^2-2u{%
u^{\prime }}^2\right]  &=&0
\end{eqnarray}
from equations (11) and (15) we get: $u=-1/3$, $u^{\prime }=0$. Which
together with $y_iy_i^{\prime }=0$, and equation (17) we can get the
following equation: 
\[
{y_v^{\prime }}^2+{y_e^{\prime }}^2+3u\left( {y_u^{\prime }}^2+{y_d^{\prime }%
}^2\right) =0\Longrightarrow Y_e^{\prime }=Y_u^{\prime },
\]
obviously it conflicts with the {\em Poincar}$\acute{e}${\em \ duality}
requirement. So these two constraints can not be satisfied at the same time
in this model. %***********************************************************

\subsection{\protect\medskip Model{\rm \ 3}:\ $H\Rightarrow M_2(C) $}

There is another model with mild change to standard model described in \cite
{ref12}, which simply changes quaternion $H\Rightarrow M_2\left( C\right) $
. Now we investigate what will happen when we input right-handed neutrinos
in it.

The representation of Algebra now is:

\[
\pi ^{+}\left( \lambda ,M_2\right) =\left( 
\begin{array}{ccc}
\overline{\lambda } & 0 & 0 \\ 
0 & \lambda & 0 \\ 
0 & 0 & M_2
\end{array}
\right) ; 
\]

We use the same bases as before with a little changing: $p_1=-1_C\oplus e$, $%
p_2=1_C\oplus s$, $p_3=1_C$. ( here the minimal-rank projection of $M_2(C)$
is not $I_2$, but $s=\left( 1,0\right) )$, and the presentation of $\gamma $
takes the same form as it in model 2.

Then the matrix $Q$:

\[
Q=\left( 
\begin{array}{ccc}
0 & -2N & -2N \\ 
-2N & 2N & 3N \\ 
-2N & 3N & 4N
\end{array}
\right) 
\]

The calculation shows $DetQ=0$. So poincar$\acute{e}$ duality requirement
alone forbids this model in NCG.

%***********************************************************

\subsection{Model {\rm 4}: $U\left( 1\right) $ extension together with $%
H\Rightarrow M_2(C)$}

Now, an idea appears naturally that whether one can combine model 2 with
model 3 to build a possible new model. We begin this work.

In the similar way as Model 3, we use the bases: $p_1=-1_C\oplus e$, $%
p_2=1_C\oplus s$, $p_3=1_C$, $p_4=1_C^{\prime }$. After carefully
calculating we get:

\[
Q=\left( 
\begin{array}{cccc}
2N\left( Y_e-Y_u\right) & -N\left( 2Y_e-Y_u\right) & -N\left( 2Y_e-Y_u\right)
& -N\left( Y_e^{\prime }-Y_u^{\prime }\right) \\ 
-N\left( 2Y_e-Y_u\right) & 2NY_e-2N & 2NY_e-N & NY_e^{\prime } \\ 
-N\left( 2Y_e-Y_u\right) & 2N\left( Y_e-1\right) & 2NY_e & NY_e^{\prime } \\ 
-N\left( Y_e^{\prime }-Y_u^{\prime }\right) & NY_e^{\prime } & NY_e^{\prime }
& 0
\end{array}
\right) 
\]

in which $Y_e=\left| y_v\right| +\left| y_e\right| $, $Y_e^{\prime }=\left|
y_v^{\prime }\right| +\left| y_e^{\prime }\right| ,$ $Y_u=\left| y_u\right|
+\left| y_d\right| $, $Y_u^{\prime }=\left| y_u^{\prime }\right| +\left|
y_d^{\prime }\right| $. The determinant is:

\[
DetQ=N^4\left( Y_e^{\prime }-Y_u^{\prime }\right) ^2 
\]

Poincar\'{e} duality condition requires: $Y_e^{\prime }-Y_u^{\prime }\neq 0$%
. As before, we next to calculate:.

\[
\pi \left( x\right) +J\pi \left( x\right) J^{\dagger }=diag\left( 
\begin{array}{c}
\left( y_v+1\right) b+y_v^{\prime }b^{\prime } \\ 
\left( y_e+1\right) b+y_e^{\prime }b^{\prime } \\ 
a+b+xb+yb^{\prime } \\ 
\left( y_u+u\right) b+\left( y_u^{\prime }+u^{\prime }\right) b^{\prime }+c
\\ 
\left( y_d+u\right) b+\left( y_d^{\prime }+u^{\prime }\right) b^{\prime }+c
\\ 
a+ub+u^{\prime }b^{\prime }+c+xb+yb^{\prime }
\end{array}
\right) 
\]

in which we use the decomposition: $u(2)=su(2)\oplus u(1)\oplus u^{\prime
}(1)$, $u(3)=su(3)\oplus u(1)\oplus u^{\prime }(1)$, then the Lie-algebra of 
$M_2(C)$ is $a+xb+yb^{\prime }$, the Lie-algebra of $M_3(C)$ is $%
c+ub+u^{\prime }b^{\prime }$ .

We get the following equations:

\begin{eqnarray}
\left( y_v+1\right) ^3+\left( y_e+1\right) ^3-2\left( x+1\right) ^3+3\left[
\left( y_u+u\right) ^3+\left( y_d+u\right) ^3-2\left( u+x\right) ^3\right]
&=&0 \\
\left( x+1\right) +3\left( u+x\right) &=&0 \\
\left( y_u+u\right) +\left( y_d+u\right) -2\left( u+x\right) &=&0 \\
\left( y_v+1\right) ^2y_v^{\prime }+\left( y_e+1\right) ^2y_e^{\prime
}-2\left( x+1\right) ^2y\ \ \ \ \ \ \ \ \ \ \ \ \ \ &&  \nonumber \\
+3\left[ \left( y_u+u\right) ^2\left( y_u^{\prime }+u^{\prime }\right)
+\left( y_d+u\right) ^2\left( y_d^{\prime }+u^{\prime }\right) -2\left(
u+x\right) ^2\left( u^{\prime }+y\right) \right] &=&0 \\
{y_v^{\prime }}^3+{y_e^{\prime }}^3-2y^3+3\left[ \left( y_u^{\prime
}+u^{\prime }\right) ^3+\left( y_d^{\prime }+u^{\prime }\right) ^3-2\left( u{%
^{\prime }+y}\right) {{}^3}\right] &=&0 \\
y+3\left( u^{\prime }+y\right) &=&0 \\
\left( y_u^{\prime }+u^{\prime }\right) +\left( y_d^{\prime }+u^{\prime
}\right) -2\left( u^{\prime }+y\right) &=&0 \\
\left( y_v+1\right) {y_v^{\prime }}^2+\left( y_e+1\right) {y_e^{\prime }}%
^2-2\left( x+1\right) y^2\ \ \ \ \ \ \ \ \ \ \ \ \ \ &&  \nonumber \\
+3\left[ \left( y_u+u\right) \left( y_u^{\prime }+u^{\prime }\right)
^2+\left( y_d+u\right) \left( y_d^{\prime }+u^{\prime }\right) ^2-2\left(
u+x\right) \left( {u^{\prime }+y}\right) ^2\right] &=&0
\end{eqnarray}

(20) and (24) tell us $2x=y_u+y_d,\ 2y=y_u^{\prime }+y_d^{\prime }$,
obviously they are integers. Together with $y_iy_i^{\prime }=0$ and (19)
(23), from equation (25) we can get :

\begin{equation}
16xy^2+{y_e^{\prime }}^2+{y_v^{\prime }}^2-\left( 4x+1\right) \left( {%
y_u^{\prime }}^2+{y_d^{\prime }}^2\right) =0
\end{equation}
which can be rewritten as: 
\[
\left( {y_e^{\prime }}^2+{y_v^{\prime }}^2\right) -\left( {y_u^{\prime }}^2+{%
y_d^{\prime }}^2\right) =-16xy^2+4x\left( {y_u^{\prime }}^2+{y_d^{\prime }}%
^2\right) =Y_e^{\prime }-Y_u^{\prime } 
\]
which is an {\bf even} integer obviously. Since $Y_e^{\prime },$ $%
Y_u^{\prime }\in \left\{ 0,1,2\right\} $, $Y_e^{\prime }-$ $Y_u^{\prime }\in
\left\{ 0,\pm 1,\pm 2\right\} $, so $Y_e^{\prime }-Y_u^{\prime }=\pm 2$ ($%
Y_e^{\prime }=Y_u^{\prime }$ violates the Poincar\'{e} duality). There are
only two cases:

Case 1: $Y_e^{\prime }=2,Y_u^{\prime }=0$

then ${y_u^{\prime }}={y_d^{\prime }=0}$, so $y=0$, put it in (26), we get $%
Y_e^{\prime }=0$, obviously it is inconsistent.

Case 2: $Y_e^{\prime }=0,Y_u^{\prime }=2$

then ${y_u}={y_d=0}$, so $x=0$, put it in (26), it obviously conflicts with
the poincar\'{e} duality requirment.

\noindent So there is no proper solution in this model.

%****************************************************************

\section{CONCLUSION}

It is unsuccessful to give each generation a massive neutrino in every model
discussed above. Maybe fundamental changes to Connes-Lott's model are
generally needed. Since majorana particles are not permitted in the present
version of NCG. Of course, if the massive neutrino contains majorana
particles, CL version of NCG should be replaced by a new one. Besides
Connes-Lott new scheme, right-handed neutrino is also considered in other
NCG versions, such as in \cite{ref15}, where the author discussed its
effects in hypercharges' determination, but unlike what we discussed in this
paper, there is no definitely constraints to judge whether the possible
existence of right-handed neutrinos is conflicted with the rigid requirements
in NCG. Above all, further experimental and theoretical researches are
needed to explore this issue. 
%************************************************************

\section{ACKNOWLEDGMENTS}

We are pleased to thank Prof. Li Chongsheng, Prof. Zheng Hanqing for
valuable discussions with them. And we also appreciate Mr. Jin Ligang, Mr.
Donghuishi and Mr. Lu Yu for giving us much help.

%\section* {REFERENCES}


\begin{references}
\bibitem{ref1}  K. Zuber, Phys. Rep. 305(1998)295.

\bibitem{ref2}  A.Connes, J.Lott, Nucl.Phys.B(Proc.Suppl)18(1990)29

\bibitem{ref3}  Connes A. $Noncommutative$ $Geometry$, Academic Press(1994)

\bibitem{ref4}  C. P. Martin, J. M. Gracia-Bondia, and J. C. Varilly, Phys.
Rep. {\bf 294}, 363(1998).

\bibitem{ref5}  Haogang Ding, Hanying Guo, Jianming Li and Ke Wu,
Z.Phys.C64(1994)521

\bibitem{ref6}  Thomas Schucker, Jean-Marec Zylinski, J. Geom. Phys. {\bf 16}
(1995) 207

\bibitem{ref7}  A.Connes, Comm.Math.Phys.182, 155-176(1996)

\bibitem{ref8}  A.Connes, J.Math.Phys. 36(1995)6194

\bibitem{ref9}  R.Brout, Nucl Phys B-Proc Sup 65: 3-15 Jun 1998

\bibitem{ref10}  Tomas Krajewski, Igor Pris,Lett Math Phys 39: (2) 187-202
Jan 1997

\bibitem{ref11}  Rich Schelp, hep-th/9905047.

\bibitem{ref12}  Igor Pris, Thomas Schucker, J Math Phys 38: (5) 2255-2265
May 1997

\bibitem{ref13}  Tomas Krajewski, J Geom Phys 28: (1-2) 1-30 Nov 1998.

\bibitem{ref14}  Ta-Pei Cheng, Ling-Fong Li. $Gauge\ theory\ of\ elementary\
particle\ physics$, Oxford University Press(1984)

\bibitem{ref15}  Jos$\acute{e}$ M. Gracia-Bond$\acute{\imath}$a, Phys .Lett
B351(1995)510
\end{references}
\end{document}